\documentclass[aps,prc,superscriptaddress,twoside,twocolumn,nofootinbib,10pt,%
showpacs,floatfix]{revtex4-1}

\usepackage{amsmath,amssymb}
\usepackage{graphicx,bm}
\usepackage{slashed}
\usepackage{epstopdf}
\usepackage{ulem} 
\usepackage[usenames]{color}
\usepackage{float}

\renewcommand\sout{\bgroup \color{red} \ULdepth=-.5ex \ULset}

\begin{document}
\title{Where is the stable Pentaquark}

\author{Woosung Park}\affiliation{Department of Physics and Institute of Physics and Applied Physics, Yonsei University, Seoul 03722, Korea}
\author{Sungtae
Cho}\affiliation{Division of
Science Education, Kangwon National University, Chuncheon 24341,
Korea}
\author{Su Houng Lee}\affiliation{Department of Physics and Institute of Physics and Applied Physics, Yonsei University, Seoul 03722, Korea}
\date{\today}
\begin{abstract}
We systematically analyze the flavor color spin structure of the
pentaquark $q^4\bar{Q}$ system in a constituent quark model based
on the chromomagnetic interaction in both the SU(3) flavor
symmetric and SU(3) flavor broken case with and without charm
quarks. We show that the originally proposed pentaquark state
$\bar{Q}s qqq$ by Gignoux et al and by Lipkin indeed belongs to
the most stable pentaquark configuration, but that when charm quark mass correction based on recent experiments are taken into
account, a doubly charmed antistrange pentaquark configuration
($udc c \bar{s}$)  is perhaps the only flavor exotic configuration
that could be stable and realistically searched for at present through the $\Lambda_c K^+ K^- \pi^+$ final states. The proposed final state is just reconstructing  $K^+$  instead of  $\pi^+$ in the measurement of $\Xi^{++}_{cc} 
\rightarrow \Lambda_c K^- \pi^+ \pi^+$ reported by LHCb collaboration 
and hence measurable immediately.
\end{abstract}


\maketitle

The possible existence of mutiquark hadrons beyond the ordinary
hadrons were first discussed for the tetraquark states in
Ref.~\cite{Jaffe:1976ig,Jaffe:1976ih} and for the H-dibaryon in
Ref.~\cite{Jaffe:1976yi}. Later, possible stable pentaquark
configurations $\bar{Q}s qqq$ were proposed in
Ref.~\cite{Gignoux:1987cn} and in Ref.~\cite{Lipkin:1987sk}. The
long experimental search for the H-dibaryon was not successful so
far but is still planned at JPARC \cite{Ahn:2017fkm}. The search
by Fermilab E791 \cite{Aitala:1999ij} for the proposed pentaquark
state also failed to find any significant signal for the exotic
configurations.

On the other hand, starting from the $X(3872)$ \cite{Choi:2003ue},
possible exotic meson configurations $XYZ$ and the pentaquark
$P_c$ \cite{Aaij:2015tga} were recently found. These states are
not flavor exotic but are known to contain $\bar{c} c$ quarks.
Heavy quarks were for many years considered to be stable color
sources that would allow for a stable multiquark configuration
that does not fall into usual hadrons. In particular, with the
recent experimental confirmation of the doubly charmed baryon
\cite{Aaij:2017ueg,Aaij:2018wzf,Aaij:2018gfl}, there is a new
excitement in the physics of exotics in general and in hiterto
unobserved flavor exotic states with  more than one heavy quarks
\cite{Chen:2017sbg,Karliner:2017qjm,Eichten:2017,Francis:2016hui,Hong:2018mpk}.

In this work, we systematically analyze the color flavor spin
structure of the pentaquark configuration within a constituent
quark model based on chromomagntic interaction. We show that the
originally proposed pentaquark state $\bar{Q}s qqq$ indeed belong
to the most stable pentaquark configuration, but that when  charm quark mass correction  based on recent
experiments are taken into
account, a doubly charmed antistrange pentaquark configuration
($\bar{s} cc ud$) is perhaps the only stable flavor exotic
configuration that could be stable and realistically searched for
at present.

{\it Systematic analysis of $q^4\bar{Q}$:} We first discuss the
classification of the flavor, color and spin wave function for the
ground state  of the pentaquark composed of $q^4$ light quarks and
one heavy antiquark $Q$ ($\bar{c}$ or $\bar{b}$) assuming that the
spatial parts of the wave function for all quarks are in the
s-wave. We categorize them into the flavor states in SU(3)$_F$,
and then examine the color $\otimes$ spin states.

The flavor states for $q^4$ can be decomposed into the direct sum
of the irreducible representation of SU(3)$_F$ as follows:
\begin{align}
[3]_F\otimes[3]_F\otimes[3]_F\otimes[3]_F=&[4]_1\otimes[15] \oplus[31]_3\otimes[{15}^\prime]  \oplus \nonumber \\
 &[21^2]_3\otimes[3] \oplus [2^2]_2\otimes[\bar{6}].
 \label{flavor}
\end{align}
Here, [4], [31], $[21^2]$ and $[2^2]$ indicate the Young tableau
of the $SU(3)_F$ multiplet, with the subscripts representing the
corresponding dimensions, while  [15], $[{15}^\prime]$, [3] and
$[\bar{6}]$ show the respective multiplicities.

The 7776 dimensional color $\otimes$ spin states of $q^4\bar{Q}$
can be classified as the  direct sum of the irreducible
 representations of SU(6)$_{CS}$ as follows:
\begin{align}
&([6]_{CS}\otimes[6]_{CS}\otimes[6]_{CS}\otimes[6]_{CS}\otimes\bar{[6]}_{CS})_{[7776]}=\nonumber \\
&[4]_1
\otimes ( [51^4] \oplus [3]
  )  \oplus
[2^2]_2
\otimes ( [3^21^3] \oplus [21]
  ) \oplus \nonumber \\
 &[31]_3
\otimes  ( [421^3] \oplus [3] \oplus [21]
  ) \oplus
[1^4]_1
\otimes ( [2^41]  \oplus [1^3]
  )\oplus  \nonumber \\
   &[21^2]_3
\otimes ( [32^21^2] \oplus [21] \oplus [1^3]
 ).
 \label{CS}
\end{align}
The Young Tableau and its subscript outside of the bracket
respectively indicates the SU(6)$_{CS}$ representation and its
dimension of the light quark sector $q^4$ while the Young Tableau
inside the bracket is the SU(6)$_{CS}$ representation of the
pentaquark consisting of $q^4\bar{Q}$.

By further decomposing the SU(6)$_{CS}$ into the sum of $SU(3)_C
\otimes SU(2)_S$ multiplets, we can select out the physically
allowed color singlet states. Table \ref{colorsinglet-spin} shows
the allowed color singlet states with the possible spin states,
denoted by $[1_C,S]$, allowed within each SU(6)$_{CS}$
representation.
\begin{table}[htp]
\caption{ The  SU(6)$_{CS}$ representations  containing $[1_C, S]$ multiplet. }
\begin{center}
\begin{tabular}{c|c}
\hline \hline
                  &    SU(6)$_{CS}$ representation    \\
\hline
$[1_C, 1/2]$  & $[2^41]$, $[32^21^2]$, [21],   $[421^3]$ \\
$[1_C, 3/2]$  &  $[1^3]$, $[32^21^2]$, $[3^21^3]$,  $[421^3]$ \\
$[1_C, 5/2]$  & $[32^21^2]$          \\
\hline \hline
\end{tabular}
\end{center}
\label{colorsinglet-spin}
\end{table}

Therefore, since the $SU(6)_{CS}$ representation of $q^4\bar{Q}$
as well as those of $q^4$ are given in Eq.~(\ref{CS}), we can
construct the  flavor $\otimes$ color $\otimes$ spin states with
color singlet,  by using the fully antisymmetric property together
with the conjugate relation between the flavor in
Eq.~(\ref{flavor}) and the $SU(6)_{CS}$ representation in
Eq.~(\ref{CS}) among the four light quarks. Such combination will
finally determine the allowed flavor and spin content of the
pentaquarks in the flavor SU(3) symmetric limit.

{\it Color spin interaction for pentaquark system:} In the
constituent quark model based on the color spin interaction, the
stability of a pentaquark depends critically on the expectation
value  of the interaction. Therefore, we derive the following
elegant  formula of the chromomagnetic interaction relevant for
the pentaquark configuration, which is similar to that of a
tetraquark in Ref. \cite{Jaffe:1976ih}, by introducing the first
kind of Casimir operator of SU(6)$_{CS}$, which is denoted by
$C^6$:
\begin{align}
- {\sum}_{i<j}^{5}\lambda_i^c\lambda_j^c{\vec{\sigma}}_i\cdot{\vec{\sigma}}_j =&
4C^6_5-8C^6_4-2C^{3}_5+4C^{3}_4-\nonumber \\
&\frac{4}{3}(\vec{S}\cdot \vec{S})_5 +\frac{8}{3} (\vec{S}\cdot  \vec{S})_4+24I.
\label{formula-2}
\end{align}
Here the lower index indicates the number of the participant
quarks, $C^3$ the first kind of Casimir operator of $SU(3)_C$, $S$
the spin operator, and $I$ the identity operator.

{\it Spin=3/2:} Let us discuss in detail the flavor
$[{15}^\prime]$ case with $S=3/2$. Here, there are two flavor
$\otimes$ color $\otimes$ spin states which are orthonormal to
each other. There are two methods to obtain these states.

In one approach based on the coupling scheme, the first (second)
state  comes from the coupling scheme of the color $\otimes$ spin
state in which the spin among the four quarks is one (two), as
given in  Eq. (26) (Eq.(32)) in~\cite{Park:2017jbn}. The fully
antisymmetric flavor $\otimes$ color $\otimes$ spin states for
$S=3/2$ among the four quarks can be obtained by  multiplying the
color $\otimes$ spin state  by their conjugate flavor
$[{15}^\prime]$ state. In the other approach, the two states can
be directly obtained from Eq.~(\ref{CS}). As we can see in
Eq.~(\ref{CS}) and Table \ref{colorsinglet-spin}, both the
$[32^21^2]$ and the $[1^3]$ $SU(6)_{CS}$ representations have the
state of the color singlet and $S=3/2$. Also, these states involve
the $[21^2]$ multiplets in the $SU(6)_{CS}$ representation among
the four quarks, which are conjugate to the flavor $[{15}^\prime]$
states, so that the two fully antisymmetric flavor $\otimes$ color
$\otimes$ spin states can be constructed from the $SU(6)_{CS}$
representation $[32^21^2]$ and $[1^3]$.

From the $SU(6)_{CS}$ representation point of view, we can infer
that the linear sum of two fully antisymmetric flavor $\otimes$
color $\otimes$ spin states coming from the coupling scheme must
belong to either $[32^21^2]$ state or $[1^3]$ state. We find that
the coefficients of the linear sum can be calculated from the
condition that these are the eigenstates of the Casimir operator
of SU(6)$_{CS}$, given by,
\begin{align}
C^6_5=&-\frac{1}{4}{\sum}_{i=1}^{4}\lambda_i^c\lambda_5^c{\vec{\sigma}}_i\cdot{\vec{\sigma}}_5
+C^6_4+\frac{1}{2}C^{3}_5-\frac{1}{2}C^{3}_4 \nonumber \\
&+\frac{1}{3}(\vec{S}\cdot \vec{S})_5-
\frac{1}{3}(\vec{S}\cdot \vec{S})_4+2I.
 \label{Casimir-6}
 \end{align}

Following the same procedure, one can construct the flavor
$\otimes$ color $\otimes$ spin states for the remaining flavor
cases for $S=3/2$, which satisfy the antisymmetry property among
four quarks. From the result, it is found that there are all
together 12 color $\otimes$ spin states that are both color
singlet and $S=3/2$. These are expressed by the Yamanouchi bases
of the $SU(6)_{CS}$ representation among the four quarks, together
with the $SU(6)_{CS}$ Young tableau for the full $q^4 \bar{Q}$
pentaquark state:
\begin{scriptsize}
\begin{align}
&\big{\vert}
 \begin{tabular}{|c|c|c|}
\hline
1 & 2 & 3     \\
\cline{1-3}
\multicolumn{1}{|c|}{4}  \\
\cline{1-1}
 \end{tabular},
 [421^3]
 \big{\rangle}
 ,
\big{\vert}
 \begin{tabular}{|c|c|c|}
\hline
1 & 2 & 4     \\
\cline{1-3}
\multicolumn{1}{|c|}{3}  \\
\cline{1-1}
 \end{tabular},
 [421^3]
 \big{\rangle}
 ,
 \big{\vert}
 \begin{tabular}{|c|c|c|}
\hline
1 & 3 & 4     \\
\cline{1-3}
\multicolumn{1}{|c|}{2}  \\
\cline{1-1}
 \end{tabular},
 [421^3]
 \big{\rangle}
 ,
 \big{\vert}
 \begin{tabular}{|c|c|}
\hline
1 & 2      \\
\cline{1-2}
\multicolumn{1}{|c|}{3}  \\
\cline{1-1}
\multicolumn{1}{|c|}{4}  \\
\cline{1-1}
 \end{tabular},
 [32^21^2]
 \big{\rangle}
 , \nonumber \\
 &\big{\vert}
 \begin{tabular}{|c|c|}
\hline
1 & 3    \\
\cline{1-2}
\multicolumn{1}{|c|}{2}  \\
\cline{1-1}
\multicolumn{1}{|c|}{4}  \\
\cline{1-1}
 \end{tabular},
 [32^21^2]
 \big{\rangle}
 ,
 \big{\vert}
 \begin{tabular}{|c|c|}
\hline
1 & 4    \\
\cline{1-2}
\multicolumn{1}{|c|}{2}  \\
\cline{1-1}
\multicolumn{1}{|c|}{3}  \\
\cline{1-1}
 \end{tabular},
 [32^21^2]
 \big{\rangle}
 ,
 \big{\vert}
 \begin{tabular}{|c|c|}
\hline
1 & 2      \\
\cline{1-2}
\multicolumn{1}{|c|}{3}  \\
\cline{1-1}
\multicolumn{1}{|c|}{4}  \\
\cline{1-1}
 \end{tabular},
 [1^3]
 \big{\rangle}
 ,
 \big{\vert}
 \begin{tabular}{|c|c|}
\hline
1 & 3    \\
\cline{1-2}
\multicolumn{1}{|c|}{2}  \\
\cline{1-1}
\multicolumn{1}{|c|}{4}  \\
\cline{1-1}
 \end{tabular},
 [1^3]
 \big{\rangle}
 , \nonumber \\
 &\big{\vert}
 \begin{tabular}{|c|c|}
\hline
1 & 4    \\
\cline{1-2}
\multicolumn{1}{|c|}{2}  \\
\cline{1-1}
\multicolumn{1}{|c|}{3}  \\
\cline{1-1}
 \end{tabular},
 [1^3]
 \big{\rangle}
 ,
 \big{\vert}
 \begin{tabular}{|c|c|}
\hline
1 & 2    \\
\hline
3 & 4    \\
\hline
 \end{tabular},
 [3^21^3]
 \big{\rangle}
 ,
 \big{\vert}
 \begin{tabular}{|c|c|}
\hline
1 & 3    \\
\hline
2 & 4    \\
\hline
 \end{tabular},
 [3^21^3]
 \big{\rangle}
 ,
\big{\vert}
 \begin{tabular}{|c|}
\hline
1 \\
\hline
2 \\
\hline
3 \\
\hline
4 \\
\hline
 \end{tabular},
 [1^3]
 \big{\rangle}
 .
  \label{12-S-3/2}
\end{align}
\end{scriptsize}

{\it Other spin states:} In analogy to the  $S=3/2$ case, we can
apply the same procedure to the  $S=1/2$ case. In this case, there
are all together 15  color $\otimes$ spin states that are both
color singlet and $S=1/2$,  and that are expressed by the
Yamanouchi bases of the $SU(6)_{CS}$ representation among the four
quarks, like Eq.~(\ref{12-S-3/2}). Finally, in the  $S=5/2$ case,
there exist only one color $\otimes$ spin state coming from the
$[32^21^2]$  representation in Table \ref{colorsinglet-spin} .

By using the flavor $\otimes$ color $\otimes$ spin states for
$S=1/2$, $S=3/2$, and $S=5/2$, the expectation values of
Eq.~(\ref{formula-2}) can be calculated, as given in Table
\ref{su3-flavor}. In Table \ref{su3-flavor}, below each matrix
element, we also show  the relevant  $SU(6)_{CS}$ representations
for the pentaquark state as well as the eigenvalue of
Eq.~(\ref{formula-2}). As can be seen in the table, the most
attractive channel is given by the $2 \times 2$ matrix valued
$(F,S)=([3],1/2)$ state. Upon diagonalizing the matrix in the $m_5
\rightarrow \infty$ one finds the eigenvalues (-16,-40/3), where
the lowest one corresponds to the most attractive pentaquark state
discussed in Ref.~\cite{Gignoux:1987cn,Lipkin:1987sk}. It should
be noted that the factor -16 in this case can also be naively
obtained by assuming two diquarks $(ud,us)$ in the $udus\bar{c}$
pentaquark. However, as noted from the case of H dibaryon, SU(3)
breaking effects together with the additional attraction from the
strong charm quarks are important to the realistic estimate of the
stability: The color spin interaction from the
$m_{J/\psi}-m_{\eta_c}$ are much stronger than from naively
scaling the color spin splitting in the light quark sector by the
charm quark mass \cite{Karliner:2017qjm}.

\begin{table}[htp]
\caption{ The expectation value of $-{\sum}_{i<j}^{5}\frac{1}{m_im_j}\lambda_i^c\lambda_j^c{\vec{\sigma}}_i\cdot{\vec{\sigma}}_j$ of $q^4\bar{Q}$ in $SU(3)_F$ limit, which means $m_4$=$m_3$=$m_2$=$m_1$. The eigenvalue indicates the value of  Eq.~(\ref{formula-2}).   }
\begin{center}
\begin{tabular}{c|c}
\hline \hline
$(F, S)$    &      $-\langle {\sum}_{i<j}^{5}\frac{1}{m_im_j}\lambda_i^c\lambda_j^c{\vec{\sigma}}_i\cdot{\vec{\sigma}}_j \rangle$  \\
 \hline
 ($[15]$, 1/2)    &   $\frac{56}{3{m_1}^2}+\frac{32}{3m_1m_5}$     \\
  $SU(6)_{CS}$    &         $[2^41]$                                          \\
 Eigenvalue      &      $\frac{88}{3}$     \\
($[15]$, 3/2)    &   $\frac{56}{3{m_1}^2}-\frac{16}{3m_1m_5}$   \\
 $SU(6)_{CS}$    &         $[1^3]$                                          \\
 Eigenvalue      &  $\frac{40}{3}$   \\
 \hline
($[{15}^{\prime}]$, 1/2)    & $\left(\begin{array}{cc}
\frac{4}{3{m_1}^2}-\frac{20}{3m_1m_5} &  \frac{4}{3{m_1}^2}+ \frac{4}{3m_1m_5}       \\
 \frac{4}{3{m_1}^2}+ \frac{4}{3m_1m_5}  &    \frac{4}{3{m_1}^2}+\frac{28}{3m_1m_5}     \\
 \end{array} \right)$        \\
  $SU(6)_{CS}$    &         $[21]$,    \qquad   \qquad  $[32^21^2]$                             \\
 Eigenvalue  &    $-\frac{8}{3}(\sqrt{10}-1)$,  $\frac{8}{3}(\sqrt{10}+1)$  \\
  ($[{15}^{\prime}]$, 3/2)    &   $\left(\begin{array}{cc}
\frac{88}{21{m_1}^2}+\frac{172}{21m_1m_5} &  \frac{16\sqrt{10}}{21{m_1}^2}+ \frac{16\sqrt{10}}{21m_1m_5}       \\
\frac{16\sqrt{10}}{21{m_1}^2}+ \frac{16\sqrt{10}}{21m_1m_5}    &    \frac{136}{21{m_1}^2}-\frac{368}{21m_1m_5}     \\
 \end{array} \right)$           \\
  $SU(6)_{CS}$    &         $[32^21^2]$,    \qquad   \qquad  $[1^3]$                             \\
 Eigenvalue  &     -12,  \qquad   \qquad \qquad $\frac{40}{3}$   \\
  ($[{15}^{\prime}]$, 5/2)    & $\frac{8}{{m_1}^2}+\frac{16}{3m_1m_5}$         \\
    $SU(6)_{CS}$    &         $[32^21^2]$                                          \\
  Eigenvalue  &  $\frac{40}{3}$         \\
 \hline
($[3]$, 1/2)    &    $\left(\begin{array}{cc}
-\frac{14}{{m_1}^2}-\frac{22}{m_1m_5} &  -\frac{2}{\sqrt{3}{m_1}^2}- \frac{2}{\sqrt{3}m_1m_5}       \\
-\frac{2}{\sqrt{3}{m_1}^2}- \frac{2}{\sqrt{3}m_1m_5}   &    -\frac{46}{3{m_1}^2}+\frac{26}{3m_1m_5}     \\
 \end{array} \right)$                      \\
   $SU(6)_{CS}$    &         $[21]$    \qquad   \qquad  $[421^3]$                             \\
    Eigenvalue  &   $-\frac{8}{3}(\sqrt{31}+8)$,   $\frac{8}{3}(\sqrt{31}-8)$    \\
($[3]$, 3/2)    &      $-\frac{40}{3{m_1}^2}+\frac{20}{3m_1m_5}$  \\
 $SU(6)_{CS}$    &         $[421^3]$                             \\
 Eigenvalue  &  $-\frac{20}{3}$         \\
\hline
($[\bar{6}]$, 1/2)    &        $-\frac{16}{3{m_1}^2}-\frac{40}{3m_1m_5}$       \\
  $SU(6)_{CS}$    &         $[21]$                             \\
 Eigenvalue  &  $-\frac{56}{3}$         \\
($[\bar{6}]$, 3/2)    &        $-\frac{16}{3{m_1}^2}+\frac{20}{3m_1m_5}$         \\
 $SU(6)_{CS}$    &         $[3^21^3]$                             \\
 Eigenvalue  &  $\frac{4}{3}$         \\
 \hline \hline
\end{tabular}
\end{center}
\label{su3-flavor}
\end{table}

{\it Pentaquark binding in the  $SU(3)_F$ broken case:} To analyze
the stability of the pentaquarks against the lowest threshold, we
introduce a simplified form for the matrix element of the
hyperfine potential term, where we approximate the spatial overlap
factors by constants that depend only on the constituent quark
masses of the two quarks involved:
\begin{align}
H_{hyp}=- {\sum}_{i<j}^{5}C_{m_im_j}\lambda_i^c\lambda_j^c{\vec{\sigma}}_i\cdot{\vec{\sigma}}_j.
\label{formula-4}
\end{align}
We then assume that the difference between the pentaquark energy
and the lowest threshold baryon meson states arise only from the
hyperfine energy difference\cite{Park:2018ukx}. This is because
other potential terms are linear in the number of quark involved
so that assuming that all hadrons occupy the same size, the
differences of their contribution to the pentaquark and baryon
meson cancel.

To evaluate the binding energy of the pentaquark in terms of
Eq.~(\ref{formula-4}), we extract the $C_{m_im_j}$ values from the
relevant mass differences between baryons and between mesons when
involving one antiquark. The relations are given by,
\begin{align}
&\Delta-P=16C_{uu}, \qquad \quad
\Sigma^*-\Sigma+\Xi^*-\Xi=32C_{us},  \nonumber \\
&\Omega^*_c-\Omega_c=16C_{sc}, \qquad
\Sigma^*_c-\Sigma_c=16C_{uc}, \nonumber \\
&2\Omega+\Delta-(2\Xi^*+\Xi)=8C_{ss}+8C_{uu}.
\label{formula-5}
\end{align}
For $C_{cc}$ we take it to be 1/2$C_{c\bar{c}}$. Then, we
calculate the binding energy, denoted by $\Delta E$, by comparing
the hyperfine potential energy between the pentaquark and its
lowest decay channel.

{\it Isopin basis:} We now investigate the stability of the
pentaquark with respect to isospin ($I$) and spin ($S$), and allow
the antiquark of the pentaquark to be either $\bar{s}$, $\bar{c}$
or $\bar{b}$.  The advantage of the Yamanuchi basis in the
$SU(6)_{CS}$ representation to the pentaquark that characterizes
the symmetry property among the four quarks makes it possible to
find the flavor $\otimes$ color $\otimes$ spin states suitable for
a certain symmetry, which is allowed by the Pauli principle. Since
it is possible that there are several flavor $\otimes$ color
$\otimes$ spin states, denoted by multiplicity, according to the
symmetry property, the binding energy is obtained from
diagonalizing the matrix element of the hyperfine potential
energy.

We need to characterize isospin states of $q^4$ in order to
classify the pentaquark with respect to $I$. As can be seen
in~\cite{Park:2016cmg}, the iospin states to $q^4$ can be
decomposed in the following way: $I=0$ with Young tableau $[2^2]$
consisting of $uudd$ component, $I=1$ with Young tableau $[31]$
consisting of $uuud$ component, and $I=2$ with Young tableau $[4]$
consisting of $uuuu$. The result for the binding energy defined as
the difference between the hyperfine interaction of the pentaquark
against its lowest threshold values are given in Table
\ref{binding}.

As can be seen in Table \ref{binding}, it is found that the most
attractive pentaquark states are those with $(I,S)=(0,1/2)$, apart
from $udcc\bar{c}$, and as well, the $uuds\bar{c}$ with
$(I,S)=(1/2,1/2)$. To understand the reason why these particles
could be bound states, we need to analyze the expectation  matrix
value of Eq.~(\ref{formula-4}) in terms of a dominant color
$\otimes$ spin state among the possible states. For these states,
the dominant color $\otimes$ spin state comes from the
$SU(6)_{CS}$ representation [21] having the Young tableau [31] for
the $q^4$\footnote{This state corresponds to the most stable P$_{\bar{c}s}$ state discussed in Ref.~\cite{Lipkin:1987sk}}, for which the expectation value of
Eq.~(\ref{formula-2}) is -36, as can be seen in $(F=[3], S=1/2)$
sector of Table \ref{su3-flavor} when $m_1=m_5$. In fact, the
SU(6)$_{CS}$ representation [21] state  with $S=1/2$ gives the
most attractive contribution to the expectation value of
Eq.~(\ref{formula-2}) than any other state, and  both the $I=0$
and $I=1/2$ comes from the breaking of the flavor [3] state of
this representation.

In Table \ref{numerical}, we show the expectation value of
Eq.~(\ref{formula-4}) in terms of only a color $\otimes$ spin
state coming from the $SU(6)_{CS}$ representation [21] as well as
the  corresponding binding energy against its threshold
represented in the third row for each state. It should be noted
that $H_{hyp}$ for each state  reduces to -36$C_{m_im_j}$ when
the $C_{m_im_j}$'s are taken to be a quark mass independent constant. It should be noted that
all these possible stable states are related to the attractive
pentaquark states discussed in Ref.
\cite{Gignoux:1987cn,Lipkin:1987sk} in the flavor SU(3) symmetric limit. However,
it should also be pointed out that when the charm quark is also
included, together with its hyperfine contribution, it is the
$udcc \bar{s}$ pentaquark configuration that is most attractive.
This state has also been discussed recently in Ref.
\cite{Zhou:2018pcv}. The attraction obtained in Table
\ref{binding} should be large enough to overcome the additional
kinetic energy, typically of order 100 MeV, to make the state
compact of a hadron size.

Hence, the proposed pentaquark state is possibly the only stable
pentaquark or a resonance state slightly above the lowest
threshold, which is $\Xi_{cc}+K$ for this state. Noting that
$\Xi_{cc}$ has been recently discovered, one can just add an
additional Kaon to look for this possible resonance state. If the
state is strongly bound, one could look at the $udc c \bar{s}
\rightarrow \Lambda_c K^+ K^- \pi^+$ decay or any hadronic decay
mode similar to those of $\Lambda_c D^+_s$.  The proposed final state is just reconstructing $K^+$ instead of  $\pi^+$ in the measurement of $\Xi^{++}_{cc} 
\rightarrow \Lambda_c K^- \pi^+ \pi^+$ reported in Ref.~\cite{Aaij:2017ueg}
and hence measurable immediately.  
 Such a measurement
would be the  first confirmation of a  flavour exotic pentaquark
state.

{\it Acknowledgement} This work was supported by the Korea
National Research Foundation under the grant number
2016R1D1A1B03930089, 2017R1D1A1B03028419(NRF) and by the National
Research Foundation of Korea (NRF) grant funded by the Korea
government (MSIP) (No. 2016R1C1B1016270).

\begin{table}[htp]
\caption{ The  expectation value of Eq.~(\ref{formula-4}) coming from the  dominant color $\otimes$ spin state for stable pentaquark candidate states.  (unit $\rm{MeV}$) }
\begin{tabular}{c|c}
\hline \hline
     $I=0$, $S=1/2$  & $udcc\bar{s}$ ($\Delta E=-131$) \\
 \hline
  $H_{hyp}$   &  11/4$C_{cc}$-11/2$C_{c\bar{s}}$-25/2$C_{uc}$-33/2$C_{u\bar{s}}$-17/4$C_{uu}$ \\
 $\Xi_{cc}K$  &   8/3$C_{cc}$-32/3$C_{uc}$-16$C_{u\bar{s}}$    \\
 \hline
     $I=0$, $S=1/2$  & $udsc\bar{c}$ ($\Delta E=-122$) \\
  \hline
  $H_{hyp}$   &  -22/3$C_{s\bar{c}}$-44/3$C_{u\bar{c}}$-28/3$C_{us}$-14/3$C_{uu}$ \\
 $\Lambda \eta_c$  &   -8$C_{uu}$-16$C_{c\bar{c}}$    \\
  \hline
     $I=0$, $S=1/2$  & $udss\bar{c}$ ($\Delta E=-113$)  \\
  \hline
  $H_{hyp}$   & 11/4$C_{ss}$-11/2$C_{s\bar{c}}$-25/2$C_{us}$-33/2$C_{u\bar{c}}$-17/4$C_{uu}$  \\
 $\Lambda D_s$  &   -8$C_{uu}$-16$C_{s\bar{c}}$    \\
   \hline
     $I=1/2$, $S=1/2$  & $uuds\bar{c}$  ($\Delta E=-92$) \\
  \hline
  $H_{hyp}$   & -33/4$C_{s\bar{c}}$-55/4$C_{u\bar{c}}$-21/2$C_{us}$-7/2$C_{uu}$  \\
 $P D_s$  &   -8$C_{uu}$-16$C_{s\bar{c}}$    \\
 \hline \hline
\end{tabular}
\label{numerical}
\end{table}

\begin{widetext}

\begin{table}[htp]
\caption{The binding energy of the pentaquark.
 (unit ; $\rm{MeV}$) }
\begin{tabular}{c|c|c|c|c|c|c|c|c}
 \hline \hline
Pentaquark     &   $I=0$,   & Threshold,  &     $I=0$,  &  Threshold,  & $I=1$,     & Threshold,
                    & $I=1$,      &   Threshold,      \\
                     &   $S=1/2$   &  multiplicity  &    $S=3/2$  &    multiplicity  &  $S=1/2$
                     & multiplicity   &   $S=3/2$    &  multiplicity      \\
\hline
$udsc\bar{c}$  &  $\Delta E$=-124  &   $\Lambda\eta_c$,   $\quad$7        &     $\Delta E$=-43  &
$\Lambda J/\psi$,  $\quad$5& $\Delta E$=-46  &  $\Sigma \eta_c$,  $\quad$8 & $\Delta E$=-31
& $\Sigma J/\psi$,  $\quad$7 \\
                    \hline
$udss\bar{c}$      &  $\Delta E$=-117  &   $\Lambda D_s$,   $\quad$4        &     $\Delta E$=-62  &
$\Lambda D^*_s$, $\quad$3  & $\Delta E$=54 &$\Sigma D_s$,$\quad$4 &$\Delta E$=1
&$\Sigma D_s^*$,$\quad$4\\
                     \hline
$udcc\bar{s}$  &  $\Delta E$=-135  &   $\Xi_{cc}K$,   $\quad$4         &     $\Delta E$=-94  &
  $\Xi^*_{cc} K$, $\quad$3 & $\Delta E$=133  & $\Xi_{cc} K$,$\quad$4 & $\Delta E$=85
   &$\Xi^*_{cc} K$,$\quad$4 \\
                     \hline
$udcc\bar{c}$  &  $\Delta E$=-38 &   $\Lambda_{c}\eta_c$,   $\quad$4        &    $\Delta E$=-43  & $\Lambda_{c}J/\psi$, $\quad$3 &$\Delta E$=14&$\Sigma_c \eta_c$,$\quad$4& $\Delta E$=-31
&$\Sigma_c^* \eta_c$,$\quad$4\\
                    \hline
$udss\bar{b}$  &  $\Delta E$=-92  &   $\Lambda B_s$,   $\quad$4         &    $\Delta E$=-67  &
$\Lambda B^*_s$,  $\quad$3  & $\Delta E$=24 & $\Sigma B_s$, $\quad$4 &$\Delta E$=20
& $\Sigma B_s^*$, $\quad$4   \\
                  \hline \hline
  \end{tabular}
 \begin{tabular}{c|c|c|c|c|c|c|c|c|c}
 Pentaquark     &   $I=0$,   & Threshold,  &     $I=0$,  &  Threshold, &  Pentaquark    & $I=1$,     & Threshold,       & $I=1$,      &   Threshold,      \\
                     &   $S=1/2$   &  multiplicity  &    $S=3/2$  &    multiplicity  &      &     $S=1/2$
                     & multiplicity   &   $S=3/2$    &  multiplicity      \\
\hline
$uudd\bar{s}$  &  $\Delta E$=98  &   $PK$,     $\quad$1      &    $\Delta E$=74
&  $PK^*$,     $\quad$1

& $uuud\bar{s}$ & $\Delta E$=337 & $PK$,$\quad$2    &$\Delta E$=-74&  $\Delta K$, $\quad$2 \\
                     \hline
$uudd\bar{c}$  &  $\Delta E$=66  &   $PD$,$\quad$1      &    $\Delta E$=58
  & $PD^*$,$\quad$1
  & $uuud\bar{c}$ & $\Delta E$=223 & $PD$, $\quad$2 &$\Delta E$=79&  $PD^*$, $\quad$2 \\
                    \hline
$uudd\bar{b}$  &  $\Delta E$=54  &   $PB$,$\quad$1      &    $\Delta E$=52  &
$PB^*$,$\quad$1
& $uuud\bar{b}$ & $\Delta E$=175 & $PB$, $\quad$2   &$\Delta E$=172 &  $PB^*$,$\quad$2  \\
                                                                \hline \hline
 \end{tabular}
 \begin{tabular}{c|c|c|c|c|c|c|c|c|c}
 Pentaquark     &   $I=1/2$,   & Threshold,  &     $I=1/2$,  &  Threshold, &  Pentaquark    & $I=1/2$,     & Threshold,       & $I=1/2$,      &   Threshold,      \\
   &   $S=1/2$   &  multiplicity  &    $S=3/2$  &    multiplicity  &      &     $S=1/2$
                     & multiplicity   &   $S=3/2$    &  multiplicity      \\
                    \hline
$uuds\bar{b}$  &  $\Delta E$=-77  &   $PB_s$,    $\quad$5       &    $\Delta E$=-45 &
$PB_s^*$,   $\quad$4
& $uuds\bar{c}$ & $\Delta E$=-99 & $PD_s$, $\quad$5 &$\Delta E$=-39
&  $P D^*_s$, $\quad$4  \\
                   \hline
$uudc\bar{s}$  &  $\Delta E$=17  &   $\Lambda_cK$,    $\quad$5   &    $\Delta E$=-88  &
$\Sigma^*_c K$ $\quad$4
& $uudc\bar{c}$ & $\Delta E$=-34 & $P\eta_c$,$\quad$5  &$\Delta E$=-15
&  $PJ/\psi$, $\quad$4 \\
                    \hline
$sssu\bar{c}$  &  $\Delta E$=133  &   $\Xi D_s$,   $\quad$3   &    $\Delta E$=-17  &
$\Xi D^*_s$,   $\quad$3
& $sssu\bar{b}$ & $\Delta E$=87 & $\Xi B_s$, $\quad$3  &$\Delta E$=73 &  $\Xi B^*_s$,$\quad$3  \\
                      \hline
 \end{tabular}
\begin{tabular}{c|c|c|c|c|c|c}
 \hline
Pentaquark     &   $I=3/2$,   & Threshold,  &     $I=3/2$,  &  Threshold,  & $I=3/2$,     & Threshold,
                      \\
                     &   $S=1/2$   &  multiplicity  &    $S=3/2$  &    multiplicity  &  $S=5/2$
                     & multiplicity        \\
\hline
$uuus\bar{c}$  &  $\Delta E$=214  &   $\Sigma D$,   $\quad$3        &     $\Delta E$=-42  &
$\Delta D_s$,  $\quad$3  & $\Delta E$=0  & $\Delta D_s^*$,  $\quad$1
 \\
   \hline
  $uuus\bar{b}$  &  $\Delta E$=170 &   $\Sigma B$,   $\quad$3        &     $\Delta E$=142  &
$\Sigma B^*$,  $\quad$3  & $\Delta E$=0  & $\Delta B_s^*$,  $\quad$1
 \\
  \hline
   $uuuc\bar{s}$  &  $\Delta E$=274 &   $\Sigma_c K$,   $\quad$3        &     $\Delta E$=186  &
$\Sigma_c^* K$,  $\quad$3  & $\Delta E$=0  & $\Delta D_s^*$,  $\quad$1
 \\
  \hline
   $uuuc\bar{c}$  &  $\Delta E$=191 &   $\Sigma_c D$,   $\quad$3        &     $\Delta E$=-20  &
$\Delta \eta_c$,  $\quad$3  & $\Delta E$=0  & $\Delta J/\psi$,  $\quad$1
 \\
  \hline
 \end{tabular}
 \begin{tabular}{c|c|c|c|c|c|c|c|c}
 \hline
 Pentaquak & $udsc\bar{c}$ & $udss\bar{c}$& $udcc\bar{s}$
 & $udcc\bar{c}$ & $udss\bar{b}$& $uuud\bar{s}$ & $uuud\bar{c}$ & $uuud\bar{b}$  \\
   \hline
 $I=1$,  $S=5/2$   & $\Delta E$=-44& $\Delta E$=-17                                                                                   &$\Delta E$=6 & $\Delta E$=0 & $\Delta E$=-17&  $\Delta E$=0   &  $\Delta E$=0   & $\Delta E$=0 \\
 Threshold, multiplicity & $\Sigma^* J/\psi$, 2  & $\Sigma^*D^*_s$, 1            & $\Xi^*_{cc}K^*$, 1   & $\Sigma^*_c J/\psi$, 1      &  $\Sigma^*B^*_s$, 1& $\Delta K^*$, 1& $\Delta D^*$, 1  &$\Delta B^*$, 1 \\
  \hline
  \end{tabular}
 \begin{tabular}{c|c|c|c|c|c}
 \hline
 Pentaquak &  $udsc\bar{c}$ &  $udss\bar{c}$& $udcc\bar{s}$
 & $udcc\bar{c}$ & $udss\bar{b}$ \\
 \hline
 $I=0$,   $S=5/2$    & $\Delta E$=-12& $\Delta E$=-7                                                                                   &$\Delta E$=-3 & $\Delta E$=-17 & $\Delta E$=-7 \\
 Threshold, multiplicity & $\Xi^*_cD^*$, 1  &  $\Xi^*D^*$, 1            & $\Xi^*_{cc}K^*$, 1    & $\Xi^*_{cc}D^*$, 1      &  $\Xi^*B^*$, 1  \\
 \hline
\end{tabular}
 \begin{tabular}{c|c|c|c|c|c|c}
 \hline
 Pentaquak &  $uuds\bar{c}$ & $uudc\bar{s}$& $uudc\bar{c}$& $uuds\bar{b}$
& $sssu\bar{c}$ & $sssu\bar{b}$  \\
 \hline
 $I=1/2$,  $S=5/2$  & $\Delta E$=-4&$\Delta E$=-1& $\Delta E$=-9                                                                                   &$\Delta E$=-4  &   $\Delta E$=-34     & $\Delta E$=-34 \\
Threshold, multiplicity &  $\Sigma^*D^*$, 1  &$\Sigma^*_cK^*$, 1  &  $\Sigma^*_cD^*$, 1       & $\Sigma^*B^*$, 1  &  $\Xi^*D^*_s$, 1     &   $\Xi^*B^*_s$, 1  \\
 \hline
  \end{tabular}
 \label{binding}
\end{table}

\end{widetext}

\end{document}